\documentclass[twocolumn,prb,amsmath,amssymb,amsfonts,superscriptaddress,floatfix,aps,showpacs,notitlepage]{revtex4-1}

\usepackage{graphicx}
\usepackage{epstopdf}
\usepackage{dcolumn}
\usepackage{bm}
\usepackage{rotating} 
\usepackage{color}
\usepackage{subfigure}
\usepackage{natbib}
\usepackage{enumitem}
\usepackage[Symbol]{upgreek}
\usepackage[symbol*]{footmisc}
\usepackage{lipsum}
\usepackage{amsmath}                    
\usepackage{amssymb}                    

\usepackage{textcomp}					

\begin{document}
\newcommand{\oscar}[1]{\textcolor{red} {#1}}
\renewcommand{\thetable}{\arabic{table}}

\title{On the origin of the quasi-particle peak in Cr(001) surfaces}
\date{\today}
\author{L. Peters}
\email{L.Peters@science.ru.nl}
\affiliation{
Institute for Molecules and Materials, Radboud University Nijmegen, NL-6525 AJ Nijmegen, The Netherlands
 }
\author{D. Jacob}
\affiliation{
Max-Planck-Institut f\"{u}r Mikrostrukturphysik, Weinberg 2, 06120 Halle, Germany
 }
\author{M. Karolak}
\affiliation{
Institut f\"{u}r Theoretische Physik und Astrophysik, Universit\"{a}t W\"{u}rzburg, Germany
 }
\author{A. I. Lichtenstein}
\affiliation{
Institut f\"{u}r Theoretische Physik, Universit\"{a}t Hamburg, Germany
 }
\author{M. I. Katsnelson}
\affiliation{
Institute for Molecules and Materials, Radboud University Nijmegen, NL-6525 AJ Nijmegen, The Netherlands
 }

\begin{abstract}
In the spectral density of Cr(001) surfaces a sharp resonance close to the Fermi level is observed in both experiment and theory. For the physical origin of this peak two mechanisms were proposed. A single particle \textit{$d_{z^{2}}$} surface state renormalised by electron-phonon coupling and an orbital Kondo effect due to the degenerate \textit{$d_{xz}$}/\textit{$d_{yz}$} states. Despite several experimental and theoretical investigations, the origin is still under debate. In this work we address this problem by two different approaches of the dynamical mean-field theory. First, by the spin-polarized T-matrix fluctuation exchange approximation suitable for weakly and moderately correlated systems. Second, by the non-crossing approximation derived in the limit of weak hybridization (i.e. for strongly correlated systems) capturing Kondo-like processes. By using recent continuous-time quantum Monte Carlo calculations as a benchmark, we find that the high-energy features, everything except the resonance, of the spectrum is captured within the spin-polarized T-matrix fluctuation exchange approximation. More precisely the particle-particle processes provide the main contribution. For the non-crossing approximation it appears that spin-polarized calculations suffer from spurious behavior at the Fermi level. Then, we turned to non spin-polarized calculations to avoid this unphysical behavior. By employing two plausible starting hybridization functions, it is observed that the characteristics of the resonance are crucially dependent on the starting point. It appears that only one of these starting hybridizations could result in an orbital Kondo resonance in the presence of a strong magnetic field like in the Cr(001) surface. It is for a future investigation to first resolve the unphysical behavior within the spin-polarized non-crossing approximation and then check for an orbital Kondo resonance.

\end{abstract}
\maketitle
\noindent

\section{Introduction}
In the growing field of spintronics the spin of the electrons is used to processes information. One popular possibility to achieve this is based on the different tunneling probabilities of spin-up and spin-down electrons in magnetic materials. Naturally these currents can be manipulated by a magnetic field. In order to apply this principle in practice for novel devices, it is crucial to understand the details behind the tunneling process~\cite{tun1,tun2,tun3}. For example, an understanding of the surface density of states of the electrodes is important. Chromium magnetic multilayers is an example where complicated many-body effects at the surface determine the tunneling~\cite{tun4}.

Besides from a technological point of view, surface science is also fundamentally interesting. New and unexpected features may occur at surfaces. An interesting example is that of the topological insulators~\cite{topin}. Another example, at the Cr(001) surface a sharp resonance close to the Fermi level is observed in angular resolved photoemission and scanning tunneling experiments~\cite{arpes1,arpes2,arpes3,dzsur1}. After this discovery many experimental and theoretical investigations were performed in order to understand the physical origin of this phenomenon. The first theoretical explanation was that of a single particle \textit{$d_{z^{2}}$} surface state~\cite{dzsur1,okon6}. However, in order to predict the correct resonance position within this picture an unrealistic reduction of the magnetic polarization was required. Based on scanning tunneling spectroscopy on very clean Cr(001) surfaces a many-body picture in terms of an orbital Kondo effect due to the degenerate \textit{$d_{xz}$} and \textit{$d_{yz}$} states was proposed~\cite{okon1,okon2}. Additional temperature dependent scanning tunneling scpectroscopy experiments followed in order to clarify the situation~\cite{okon3}. However, it appeared that both models were in agreement with the experimental data. Although for the \textit{$d_{z^{2}}$} single particle picture an electron-phonon coupling strength 5-10 times larger than in the bulk was required. By combining scanning tunneling microscopy, photoemission spectroscopy and inverse photoemission spectroscopy one was able to show that the resonance above the Fermi level was mainly of \textit{$d_{z^{2}}$} character~\cite{okon4}. This contradicts the orbital Kondo interpretation. Although, one should realise that the resolution of inverse photoemission spectroscopy is too low to properly investigate the sharp resonance at low temperatures. On the other hand for the \textit{$d_{z^{2}}$} single particle picture the large electron-phonon enhancement compared to the bulk remains questionable. The newest photoemission experiments show a pseudogap below roughly 200~K and the emergence of a sharp resonance below 75~K~\cite{okon5}. Note that this type of behavior was not observed in earlier experiments. These newest experiments hint in the direction of a many-body interpretation of the resonance. Also recent dynamical mean-field theory (DMFT) calculations within the continuous-time quantum Monte Carlo (CTQMC) solver hint in this direction~\cite{schuler}. Namely it was observed that the resonance was very robust against artificial shifts in the one-particle energies of the \textit{$d_{xz}$}, \textit{$d_{yz}$} and \textit{$d_{z^{2}}$} states, which points towards a dominant many-body contribution. 

There are several disadvantages involved with employing the CTQMC solver. For example, it is very difficult to access temperatures at which the resonance is observed experimentally. Further, the consideration of the full Coulomb matrix becomes prohibitively expensive within CTQMC at low temperatures. Therefore, in Ref.~\onlinecite{schuler} the lowest temperature that could be considered was still too high to observe the sharp resonance and only the density-density terms of the Coulomb matrix were taken into account. It is known that such an approximation to the Coulomb matrix can lead to qualitatively wrong results.~\cite{fullcoul1,fullcoul2} Apart from the approximation in the Coulomb matrix, the CTQMC method is essentially exact, i.e. all Feynmann diagrams are taken into account. Therefore, it is very hard to obtain a detailed understanding of the physical processes responsible for the observed spectral features. In order to avoid these disadvantages of the CTQMC method, we employed two approximate methods derived in two opposite limits and able to consider the full Coulomb matrix for temperatures far below where the sharp resonance is observed. The spin-polarized T-matrix fluctuation exchange (SPTF) approximation is derived in the limit of weak and moderate correlations in which the interaction can be treated perturbatively.~\cite{sptf1,sptf2} Although SPTF is known not to capture Kondo-like physics properly, it can be used to test whether the resonance has some other many-body origin. The non-crossing approximation (NCA) is derived in the limit of strong correlations, where the hybridization is treated as a perturbation.~\cite{nca1,nca2} Note that the NCA is basically designed to capture (orbital and spin) Kondo-like processes and is therefore the ideal candidate to test for the orbital Kondo effect~\cite{davidprl}.

From the limits in which SPTF and NCA are derived, it is clear that both methods consider totally different physical processes. By using the recent CTQMC results as a benchmark, we are able to trace down the physical processes responsible for the high-energy spectral features, everything except the resonance. These are the particle-particle processes within SPTF. For the NCA it appeared that spin-polarized calculations suffer from spurious behavior at the Fermi level. Then, we turned to non spin-polarized calculations to avoid this unphysical behavior. By employing two plausible starting hybridization functions, it is observed that the characteristics of the resonance are crucially dependent on the starting point. It appears that only one of these starting hybridizations could result in an orbital Kondo resonance in the presence of a strong magnetic field like in the Cr(001) surface. However, to unambiguously establish this, first a thorough investigation is required in order to resolve the unphysical behavior at the Fermi level within the spin-polarized NCA. Such an investigation is out of the scope of this work.

In the following we first give a description of the SPTF and NCA methods. Then, we discuss the results of these methods and finally we make a conclusion.

\section{Theory}
 \subsection{Dynamical mean-field theory}
Density functional theory (DFT) in its conventional local density approximation (LDA) or generalized gradient approximation (GGA) is known to be quite successful in predicting properties of real materials, i.e. structural properties, magnetic moments and band structures~\cite{dft1,dft2,lda1,lda2,gga}. Since DFT is essentially a single particle approximation, and  LDA and GGA are derived in the limit of a (nearly) uniform electron gas, this usually only holds for weakly correlated systems. For moderately and strongly correlated systems a proper treatment of correlation effects is missing. However, even for weakly correlated systems DFT will never be able to capture pure many-body effects like quasi-particle life-times or resonances.

At that time it was also realized that Hubbard-like models perform well in describing (strong) correlation effects, i.e. Mott-insulator transition and quasi-particle peaks. Therefore, the idea came to describe the delocalized weakly correlated electrons of a system within DFT and for the strongly correlated electrons add by hand the most important missing part. From experience with Hubbard-like models this missing part is the onsite Coulomb interaction. Thus, this leads to a generalized Hubbard model.

The main problem is to accurately solve this generalized Hubbard model for all interaction strengths. A huge breakthrough came with the discovery of the dynamical mean-field theory~\cite{dmft1,dmft2}. It was shown that in the limit of infinite dimensions or equivalently infinite nearest neighbors the self-energy becomes purely local. In other words in this limit only local diagrams survive leading to a k-independent self-energy. Since the topology of these diagrams are the same as those of an Anderson impurity model, the generalized Hubbard model can be mapped onto this model. The great advantage of this is that for the Anderson impurity model solvers exist. Thus, by peforming a mapping to the Anderson impurity model and then using one of the solvers, the local self-energy of the lattice problem (generalized Hubbard model) is obtained. This is a good approximation when the self-energy is purely local, i.e. in the limit of infinite nearest neighbors. However, from experience it is known that this limit is reached rather fast, already for two or three dimensions.

By using quantum Monte Carlo methods, for example continuous-time quantum Monte Carlo (CTQMC), the Anderson impurity model can be solved numerically exactly, i.e. all Feynmann diagrams are taken into account~\cite{qmc1,qmc2}. However, the quantum Monte Carlo methods also have several disadvantages. One of them is that low temperatures are very hard to access. Another, the consideration of the full Coulomb matrix becomes prohibitively expensive at low temperatures. Further, since all diagrams are considered, it becomes very hard to obtain a detailed understanding of which physical processes are responsible for the observed spectral features. To avoid these disadvantages, numerically efficient perturbative solvers have been developed that are able to consider the full Coulomb matrix at low temperatures. In the limit of weak or moderate correlations the iterative perturbation theory and spin-polarized T-matrix fluctuation exchange approach have been derived~\cite{ipt,sptf1,sptf2}. For the limit of strong correlations, where the hybridization can be treated perturbatively, the non-crossing and one-crossing approximation have been formulated~\cite{nca1,nca2,oca}.

 \subsection{SPTF}
The idea of SPTF is to find a numerically efficient approach for the Anderson impurity model in the limit of weak (and moderate) correlations. In order to achieve this, the interaction is treated perturbatively. More precisely, diagrams known to be dominant for systems with low electron densities (and short-range repulsive potential) and high electron densities are considered. Since SPTF is exact in these two limits, it is also thought to provide an accurate description for systems with intermediate densities. From a large variety of SPTF calculations, it has appeared that a qualitatively satisfactory description of weak and moderate correlated systems can be obtained.~\cite{sptfig1,sptfig2}.

The dominant diagrammatic contribution for a low density electron system with short-ranged repulsive interaction comes from the ladder diagrams in the particle-particle channel. The particle-particle channel consists of electron-electron and hole-hole contributions. It can be shown that in the regime of low densities the former dominates the latter. The particle-particle contribution to the self-energy within SPTF is given by

\begin{equation}
\begin{gathered}
\Sigma_{m_{1},m_{2}}^{TH}(i\omega_{n})=\\
\frac{1}{\beta}\sum_{i\Omega_{m}}\sum_{m_{3},m_{4}}T_{m_{1},m_{3},m_{2},m_{4}}(i\Omega_{m})G_{m_{4},m_{3}}(i\Omega_{m}-i\omega_{n})\\
\Sigma_{m_{1},m_{2}}^{TF}(i\omega_{n})=\\
\frac{1}{\beta}\sum_{i\Omega_{m}}\sum_{m_{3},m_{4}}T_{m_{1},m_{4},m_{3},m_{2}}(i\Omega_{m})G_{m_{3},m_{4}}(i\Omega_{m}-i\omega_{n}).
\end{gathered}
\label{eqsptfpp}
\end{equation}
\newline
Here $\beta$ is the inverse temperature, $G$ is the single particle Green's function, the $m_{x}$ labels refer to the strongly correlated orbitals, $\Omega$ and $\omega$ are respectively bosonic and fermionic Matsubara frequencies. Further, $\Sigma_{m_{1},m_{2}}^{TH}$ and $\Sigma_{m_{1},m_{2}}^{TF}$ correspond to the Hartree and Fock contributions with an effective interaction defined in terms of the T-matrix

\begin{equation}
T(i\Omega_{m})=U-U\star \chi^{PP}(i\Omega_{m})\star T(i\Omega_{m}).
\label{Tmatrix}
\end{equation}
\newline
This equation is in terms of 4 index matrices, where $\star$ represents the according matrix multiplication. The $U$ represents here the bare onsite Coulomb interaction and $\chi^{PP}$ has a convenient representation in imaginary time

\begin{equation}
\chi^{PP}_{m_{1},m_{2},m_{3},m_{4}}(\tau)=G_{m_{1},m_{3}}(\tau)G_{m_{2},m_{4}}(\tau).
\label{chipp}
\end{equation}
 \newline
Note that the contributions of Eq.~\ref{eqsptfpp} include all first and second order contributions in the bare interaction exactly.

In the high density electron limit the electron-hole bubble contributions become dominant, the random phase approximation. Besides this contribution there is another term known to be important for the description of magnetic fluctuations, the particle-hole ladder contribution. Both particle-hole contributions can be conveniently taken into account by introducing the following anti-symmetrized vertex

\begin{equation}
U^{AS}_{m_{1},m_{2},m_{3},m_{4}}=T_{m_{1},m_{2},m_{3},m_{4}}(0)-T_{m_{1},m_{2},m_{4},m_{3}}(0).
\label{asv}
\end{equation}
\newline
Here the bare interaction has been replaced by the static value of the T-matrix of Eq.~\ref{Tmatrix}, because these ladder particle-particle processes are known to be important for the renormalization of the interaction~\cite{sptf1}. Then, the particle-hole contribution to the self-energy within SPTF can be written as

\begin{equation}
\Sigma_{m_{1},m_{2}}^{PH}(\tau)=\sum_{m_{3},m_{4}}W_{m_{1},m_{3},m_{4},m_{2}}(\tau)G_{m_{4},m_{3}}(\tau).
\label{eqsptfph}
\end{equation}
\newline
Here the particle-hole fluctuation potential is given by

\begin{equation}
W(\Omega)=U^{AS} \star \chi^{PH}(i\Omega)\star \big[I-U^{AS} \star \chi^{PH}(i\Omega)\big]^{-1} \star U^{AS} - W_{2}(i\Omega),
\label{wph}
\end{equation}
\newline
where the particle-hole susceptibility is

\begin{equation}
\chi^{PH}_{m_{1},m_{2},m_{3},m_{4}}(\tau)=-G_{m_{4},m_{1}}(-\tau)G_{m_{2},m_{3}}(\tau).
\label{chiph}
\end{equation}
\newline
The term $W_{2}$ in Eq.~\ref{wph} is required to remove the second order contribution, which is already contained in Eq.~\ref{eqsptfpp}.

\subsection{NCA}
The NCA is a numerically efficient solver for the Anderson impurity model derived in the limit of strong correlations. In this limit the hybridization can be treated perturbatively. However, the machinery of quantum field theory (Wick's theorem) cannot be applied straightforwardly, because the zeroth order term contains the many-body onsite interaction term explicitly. The zeroth order term is given by $H_{imp}$ in the (multiple orbital) Anderson impurity model

\begin{equation}
\begin{gathered}
H=H_{imp}+H_{bath}+V_{hyb}\\
H_{imp}=\sum_{\alpha,\sigma}\epsilon_{\alpha\sigma}d_{\alpha\sigma}^{\dagger}d_{\alpha\sigma}+\\
\frac{1}{2}\sum_{\alpha,\beta,\alpha',\beta',\sigma,\sigma'}U_{\alpha\alpha'\beta\beta'}d_{\alpha\sigma}^{\dagger}d_{\alpha'\sigma'}^{\dagger}d_{\beta'\sigma'}d_{\beta\sigma}\\
H_{bath}=\sum_{k,\nu,\sigma}\epsilon_{k\nu\sigma}c_{k\nu\sigma}^{\dagger}c_{k\nu\sigma}\\
V_{hyb}=\sum_{k,\nu,\sigma}V_{k\nu,\alpha}\Big(d_{\alpha\sigma}^{\dagger}c_{k\nu\sigma}+c_{k\nu\sigma}^{\dagger}d_{\alpha\sigma}\Big).
\label{eqAIM}
\end{gathered}
\end{equation}
\newline
Here $\epsilon_{\alpha\sigma}$ are the single-particle impurity energy levels and $U_{\alpha\alpha'\beta\beta'}$ is the onsite Coulomb repulsion between the impurity states. Further, $H_{bath}$ resprents the bath of non-interacting electrons whose dispersion is given by $\epsilon_{k\nu\sigma}$. The last term $V_{hyb}$ describes the coupling between the impurity and bath states.

 By rewriting Eq.~\ref{eqAIM} in terms of pseudo-particles, the standard field theoretical perturbation theory can be employed again. Each pseudo-particle corresponds to a many-body eigenstate $|m\rangle$ and eigenenergy $E_{m}$ of the isolated impurity

\begin{equation}
H_{imp}=\sum_{m}E_{m}|m\rangle\langle m|.
\label{eqisimp}
\end{equation}
\newline
Based on these eigenstates $|m\rangle$, pseudo-particle creation $a_{n}^{\dagger}$ and annihilation $a_{m}$ operators can be introduced with the following relation to the physical electron operators

\begin{equation}
d_{\alpha\sigma}=\sum_{n,m}F_{nm}^{\alpha\sigma}a_{n}^{\dagger}a_{m}.
\label{relphpp}
\end{equation}
\newline
Here $F_{nm}^{\alpha\sigma}=\langle n|d_{\alpha\sigma}|m\rangle$ is the matrix element of the physical impurity electron operator. In terms of the pseudo-particle operators the Anderson impurity model is written as

\begin{equation}
\begin{gathered}
H=\sum_{m}E_{m}a_{m}^{\dagger}a_{m}+\sum_{k\nu\sigma}\epsilon_{k\nu}c_{k\nu\sigma}^{\dagger}c_{k\nu\sigma}\\
+\sum_{m,n,k,\nu,\alpha,\sigma}\big(V_{k\nu,\alpha}F_{nm}^{\alpha\sigma}c_{k\nu\sigma}^{\dagger}a_{m}^{\dagger}a_{n}+h.c.\big).
\label{eqAIMpp}
\end{gathered}
\end{equation}
\newline
From this expression it is clear that the field theoretical perturbative techniques can be employed again, where the hybridization is now the interaction term. It describes the interaction among the pseudo-particles induced by the coupling to the bath electrons. Thus, the problem is to find a good approximation for the pseudo-particle self-energy $\Sigma_{m}(\omega)$ of the pseudo-particle propagator

\begin{equation}
G_{m}(\omega)=\big(\omega-\lambda-E_{m}-\Sigma_{m}(\omega)\big)^{-1}.
\label{greenpp}
\end{equation}
\newline
Here $\lambda$ is the Lagrange multiplier of the Lagrangian constraint $\lambda(Q-1)$, which is required to ensure the completeness of the impurity eigenstates

\begin{equation}
Q=\sum_{m}a_{m}^{\dagger}a_{m}=1.
\label{constr}
\end{equation}
\newline
Within NCA the pseudo-particle self-energy is approximated by an infinite resummation of diagrams with non-crossing conduction electron lines, which is exact to first order in the hybridization function

\begin{equation}
\Delta_{\alpha}(\omega)=\sum_{k,\nu}V_{k\nu,\alpha}^{\ast}g_{k\nu}V_{k\nu,\alpha}.
\label{ncahybfun}
\end{equation}
\newline
Here $g_{k\nu}(\omega)=(\omega^{+}+\mu-\epsilon_{k\nu})^{-1}$ is the bare bath electron propagator. The diagrams included in NCA describe the processes where a single impurity electron (hole) hops to the bath and back. Hereby a pseudo-particle with $N+1$ ($N-1$) electrons is temporarily created. Notice that these processes are known to be responsible for the appearance of the Kondo peak at low enough temperatures. For completeness the expression for the pseudo-particle self-energy in NCA is given by

\begin{equation}
\begin{gathered}
\Sigma_{m}^{NCA}(\omega)=-\sum_{m',\alpha,\sigma}\Big[|F_{mm'}^{\alpha\sigma}|^{2}\int\frac{dv}{\pi}f(\nu)\Delta_{\alpha}^{"}(\nu)G_{m'}(\omega+\nu)\\
+|F_{m'm}^{\alpha\sigma}|^{2}\int\frac{dv}{\pi}f(-\nu)\Delta_{\alpha}^{"}(\nu)G_{m'}(\omega-\nu)\Big].
\label{selfnca}
\end{gathered}
\end{equation}
\newline
Here $\Delta_{\alpha}^{"}(\nu)$ is the imaginary part of the hybridization function in Eq.~\ref{ncahybfun} and $f(\nu)$ is the Fermi function. After the pseudo-particle self-energies are obtained self-consistently, they need to be translated in order to obtain real physical quantities.

NCA is known to provide a good qualitative description of the Kondo resonance and Hubbard subbands~\cite{nca1,nca2}. Shortcommings are an underestimation of the Kondo temperature, an overestimation of the asymmetry and height of the Kondo resonance, and for temperatures much smaller than the Kondo temperature a spurious peak emerges at the Fermi level due to missing vertex corrections~\cite{costi,kura}.


\subsection{Computational details}
The DFT(+SPTF) calculations reported here were carried out using a full potential linear muffin-tin orbital (FP-LMTO) method~\cite{RSPt_book}. The GGA parametrization of Perdew, Burke, and Ernzerhof was used~\cite{gga}. The Brillouin zone was sampled through a conventional Monkhorst-Pack mesh of 20 x 20 x 1 {\bf{k}}-points, leading to a total of 102 vectors in the irreducible wedge. The basis setup was the same for all calculations. For the definition of the muffin-tin sphere of Cr a radius of 2.23~a.u. is used. The main valence basis functions were chosen as 3d, 4s and 4p states, while 3s and 3p electrons were treated as core states~\cite{RSPt_book}. Three kinetic energy tails were used for 4s and 4p states, corresponding to the default values 0.3, -2.3 and -0.6 Ry. Only the first tail is used for the 3d states. The use of a single tail is due to the choice of the construction of the correlated orbitals of the Anderson impurity model. These correlated orbitals are constructed from LMTOs, that have a representation involving structure constants, spherical harmonics, and a numerical radial representation inside the muffin-tin spheres. These functions are matched continuously and differentiably at the border of the muffin-tin spheres to Hankel or Neumann functions in the interstitial. The "ORT" basis originates from these native LMTOs after a L\"owdin orthonormalization. The MT orbitals, instead, are atomic-like orbitals where the radial part comes from the solution of the radial Schr\"odinger equation inside the muffin-tin sphere at an energy corresponding to the 'center of gravity' of the relevant energy band. For a more detailed description of the correlated orbital bases we refer to Ref.~\onlinecite{sptfig1}. There, it is also shown that they generally lead to very similar results. In this work the ORT basis is used.

As for the double counting within DFT+SPTF the orbitally averaged static part of the self-energy is used. For the parameterization of the onsite Coulomb interaction the constrained random phase approximation results of Ref.~\onlinecite{ersoy} are used. In their work a slab of 10 layers is considered for which they found the following onsite Coulomb interactions $U_{1/10}=3.44$~eV, $U_{2/9}=4.64$~eV, $U_{3/8}=4.73$~eV, $U_{4/7}=4.94$~eV and $U_{5/6}=4.95$~eV. Here the numbers indicate the layer of the slab, i.e. 1 and 10 are respectively the top and bottom layer. The Hund exchange interaction is constant $J=0.65$~eV.

In order to use the CTQMC results as a benchmark the Cr(001) surface is modelled in exactly the same way as in Ref.~\onlinecite{schuler}. This is a slab of 10 atomic layers with a vacuum of approximately 16~\AA$ $ stacked in the z-direction and perdiodically continued in the x and y direction. This structure is optimized by allowing the atomic coordinates to relax in the z-direction. Further, we also break the symmetry in an anti-ferromagnetic way in the first iteration of the DMFT loop. Note that we also started the spin-polarized DMFT calculation on top of a converged non spin-polarized DFT (GGA) calculation. Just as the CTQMC calculations our calculations are not charge self-consistent. Also in accordance with the CTQMC calculations, is the application of a multi-site version of the DMFT method to model the slab of 10 atomic layers. On the Matsubara axis the lattice Green's function within the multi-site version of DMFT is given by

\begin{equation}
\begin{gathered}
G_{\alpha \beta}^{ij}(i\omega_{n},{\bf k})^{-1}=\\
[ (i\omega_{n}+\mu)\delta_{\alpha \beta}- \Sigma_{\alpha \beta}^{i}(i\omega_{n}) ]\delta_{ij} - H({\bf k})_{\alpha \beta}^{ij}.
\label{muldmft}
\end{gathered}
\end{equation}
\newline
Here $i$ and $\alpha$ refer to the local basis functions $|i,\alpha \rangle$ with $i$ and $\alpha$ respectively corresponding to the site and correlated orbital. The chemical potential is represented by $\mu$ and for completeness $ H({\bf k})_{\alpha \beta}^{ij}=\langle i,\alpha|H({\bf k})|j,\beta \rangle$. Further, $\Sigma_{\alpha \beta}^{i}(i\omega_{n})$ is the local self-energy, i.e. it is ${\bf k}$-independent and $i\neq j$ terms are zero. The double-counting correction is absorbed in the self-energy. In order to obtain the rest of the self-energy, an effective impurity model is solved for each Cr atom in the slab until self-consistency within the DMFT loop is reached. For this purpose at each DMFT iteration the following site dependent Weiss fields are computed
 
\begin{equation}
\mathcal{G}_{0,\alpha \beta}^{i}(i\omega_{n})^{-1}= G_{\alpha \beta}^{i}(i\omega_{n})^{-1} + \Sigma_{\alpha \beta}^{i}(i\omega_{n}).
\label{muldmft2}
\end{equation}
\newline  
Here $G(i\omega_{n})$ is the local lattice Green's function, which is obtained by taking a ${\bf k}$-average of $G(i\omega_{n},{\bf k})$. 

For the NCA calculations we did not use the multi-site DMFT version, since we performed only one-shot DMFT calculations. Then, the standard single-site version can be employed.  For a one-shot NCA calculation we obtain the hybridization function and projected 3d-eigenvalues from a converged non spin-polarized GGA calculation. This GGA calculation is performed for the same geometry as described above. The double counting is about 13.5~eV in order to have a total of approximately 4.75 3d-electrons.

\section{Results}

\subsection{GGA}
\begin{figure}[!ht]
\begin{center}
\includegraphics[trim=80 40 30 60, clip, width=9cm, scale=0.5]{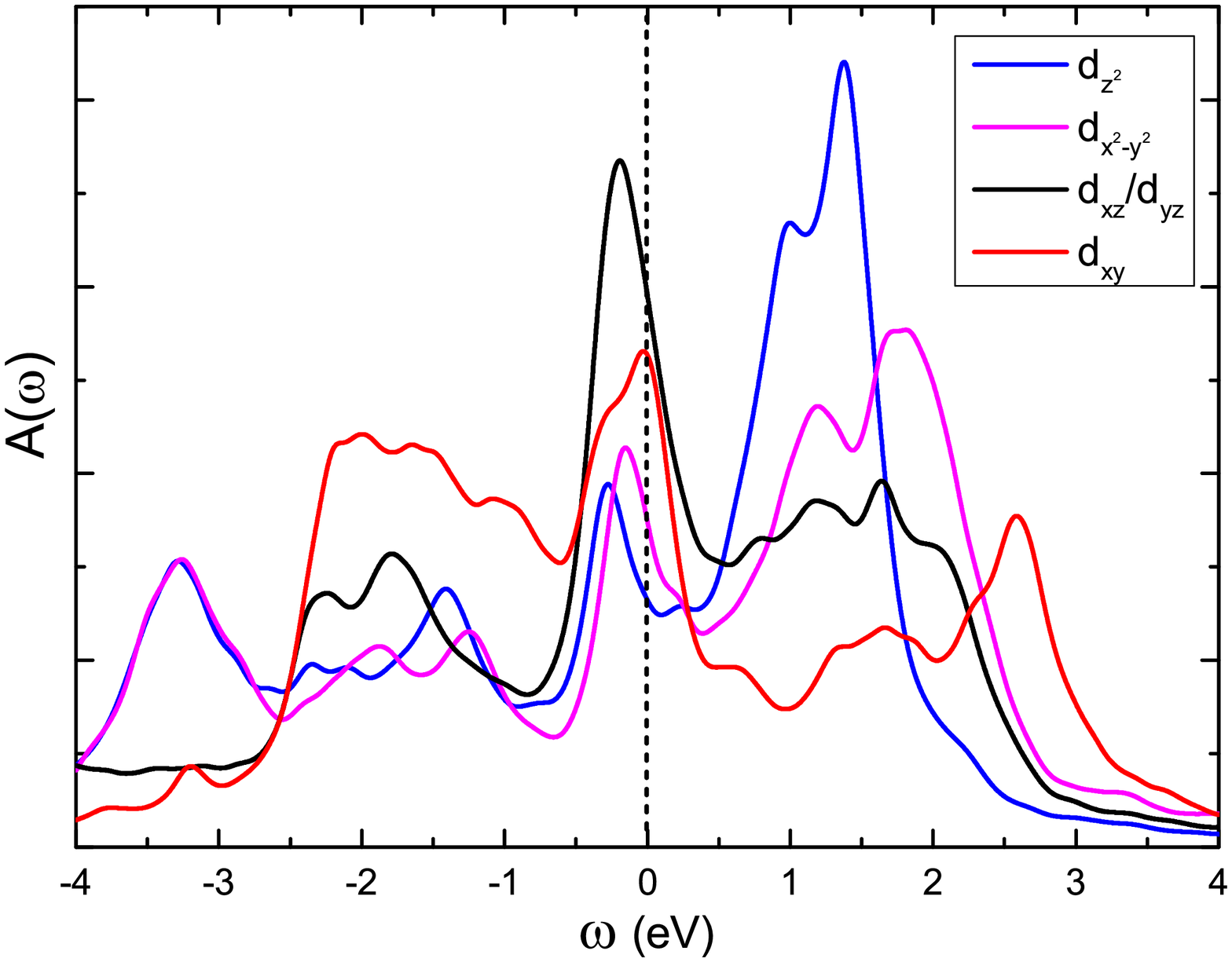}
\includegraphics[trim=80 40 30 60, clip, width=9cm, scale=0.5]{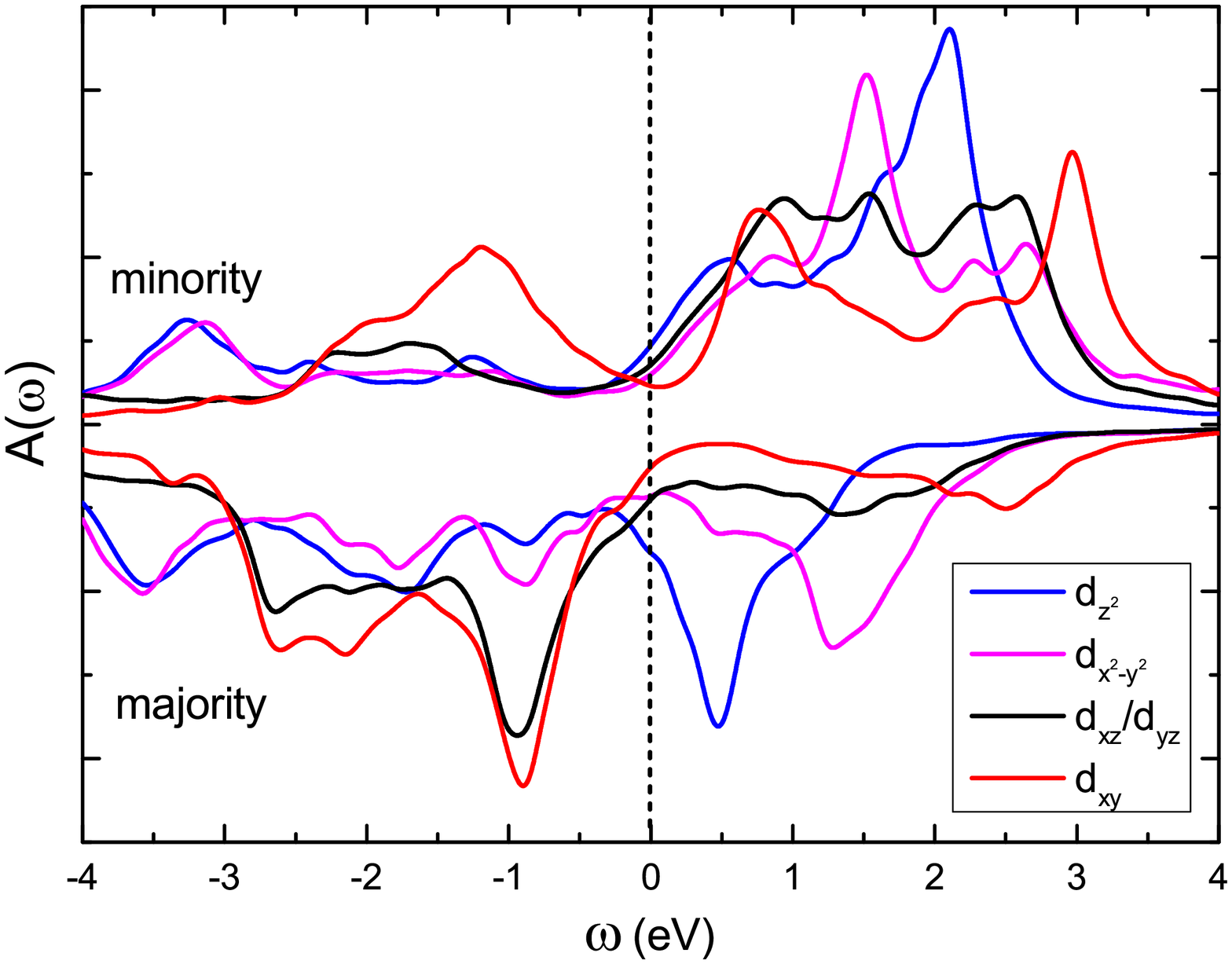}
\end{center}
\caption{The 3d projected partial density of states is plotted for a non spin-polarized (top) and a spin-polarized GGA calculation (bottom). Here blue corresponds to \textit{$d_{z^{2}}$}, magenta to \textit{$d_{x^{2}-y^{2}}$}, black to \textit{$d_{xz}$}/\textit{$d_{yz}$} and red to \textit{$d_{xy}$}. }
\label{ggafig}
\end{figure}

Before we study the many-body effects within DMFT on the spectral properties, we first consider the single-particle GGA approach. In Fig.~\ref{ggafig} the projected density of states of the 3d-states are plotted for a non spin-polarized and spin-polarized GGA calculation. These spectra are in very good agreement with what is reported in literature~\cite{okon2,okon6,schuler}. For example the results of Ref.~\onlinecite{schuler}, for convenience presented here in Fig.~\ref{CTQMCfig}a and b, are very similar to our results in Fig.~\ref{ggafig}. The non spin-polarized calculation is convenient to make a rough estimate of the bandwidth. From the top figure of Fig.~\ref{ggafig} it can be observed that the bandwidth is about 7~eV. As mentioned above from constrained RPA calculations it is known that for the top surface layer the screened onsite correlations within these 3d-states is 3.44~eV. This suggests that the 3d-states of the Cr(001) surface are weakly/moderately correlated.

From the bottom figure of Fig.~\ref{ggafig} it is clear that inclusion of spin polarization has a huge effect on the spectral density. Further, the exchange induced spin splitting can be observed. By comparing Fig.~\ref{ggafig} (both non spin-polarized and spin-polarized) with experiment~\cite{arpes1,arpes2,arpes3,okon1,okon2,okon3,okon4}, it can be concluded that GGA is not able to account for the resonance. For the non spin-polarized case the peaks at the Fermi level are too high and broad, and the orbital character is not in accordance with experiment~\cite{okon4}. On the other hand for the spin-polarized case there is no peak at the Fermi level.


\subsection{SPTF}
\begin{figure}[!ht]
\begin{center}
\includegraphics[trim=320 210 50 330, clip, width=9cm, scale=0.5]{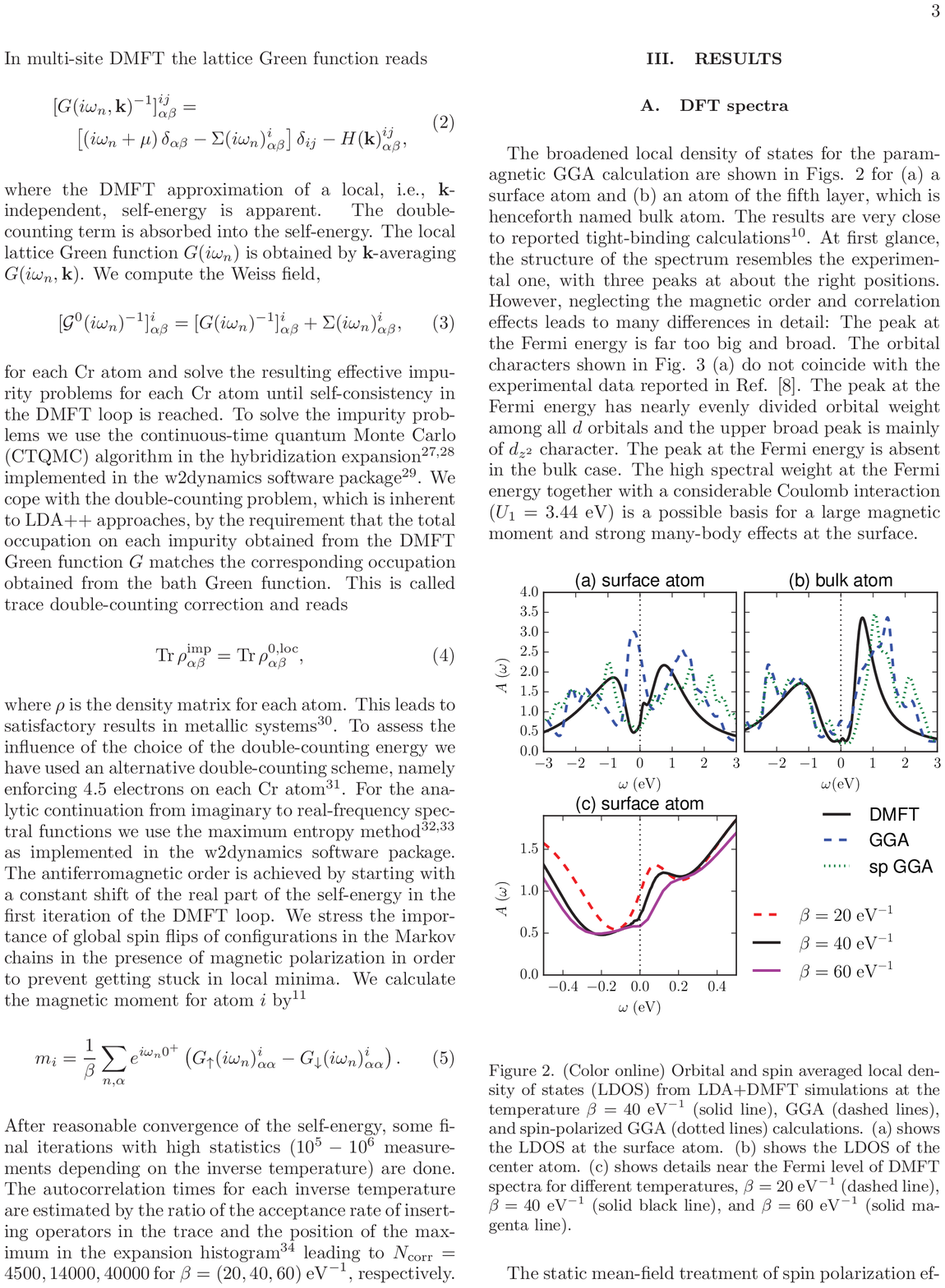}
\end{center}
\caption{The CTQMC local spin averaged 3d partial density of states for different methods and inverse temperatures, $\beta=20$~eV$^{-1}$ (dashed red), $\beta=40$~eV$^{-1}$ (solid black) and $\beta=60$~eV$^{-1}$ (solid magenta) of Ref.~\onlinecite{schuler}.  }
\label{CTQMCfig2}
\end{figure}

\begin{figure}[!ht]
\begin{center}
\includegraphics[trim=58 295 310 285, clip, width=9cm, scale=0.5]{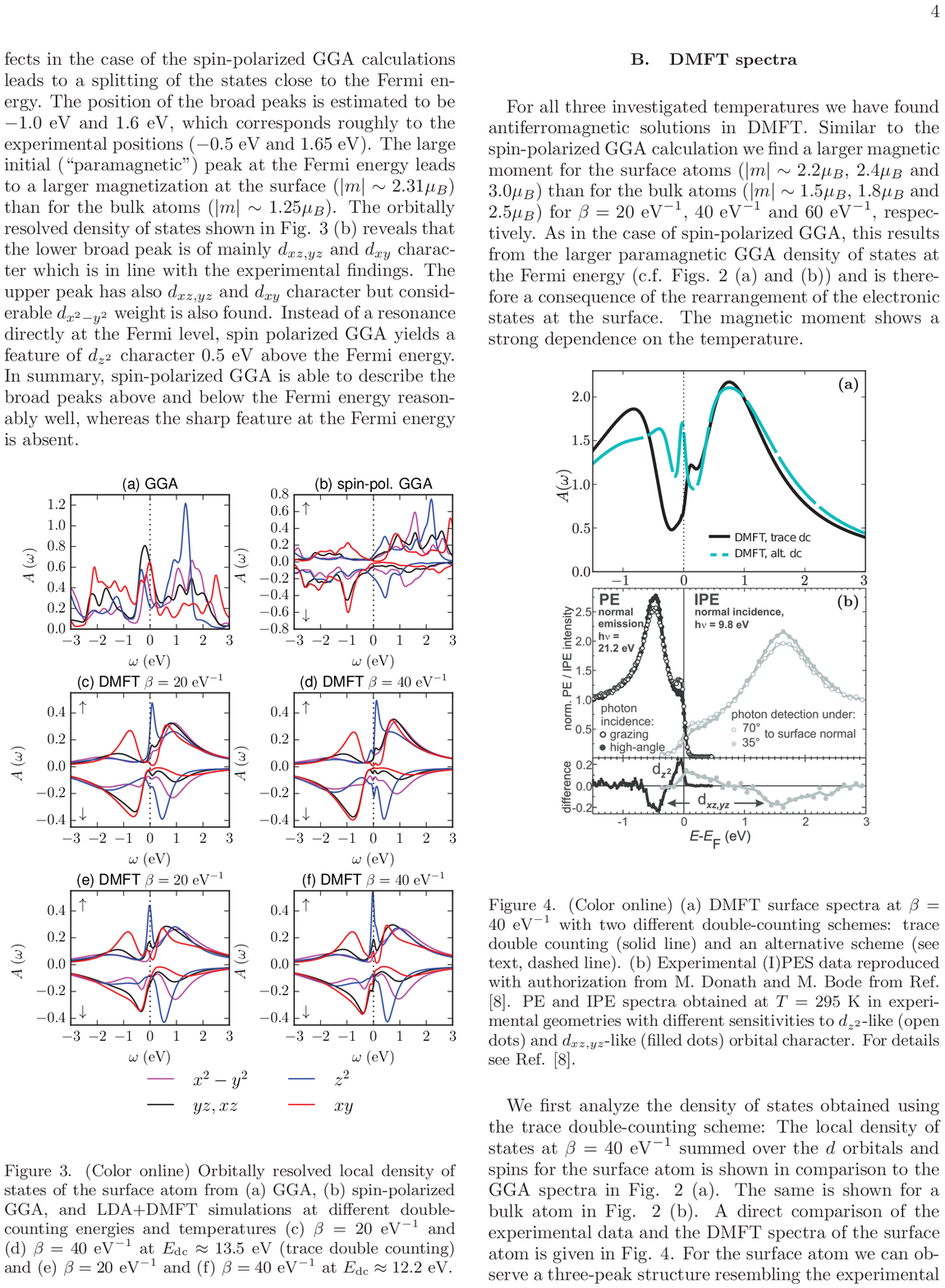}
\end{center}
\caption{The CTQMC 3d projected partial density of states of Ref.~\onlinecite{schuler}. Here the top two figures are for non spin-polarized and spin-polarized GGA and the bottom figures for CTQMC at two different inverse temperatures, $\beta=20$~eV$^{-1}$ (c), $\beta=40$~eV$^{-1}$ (d). Here blue corresponds to \textit{$d_{z^{2}}$}, magenta to \textit{$d_{x^{2}-y^{2}}$}, black to \textit{$d_{xz}$}/\textit{$d_{yz}$} and red to \textit{$d_{xy}$}. }
\label{CTQMCfig}
\end{figure}

Since CTQMC is in principle exact we use it as a benchmark for our approximate solvers. More precisely, we compare our results with those of Ref.~\onlinecite{schuler} presented here in Figs.~\ref{CTQMCfig2} and \ref{CTQMCfig}. In the former the local spin averaged 3d partial density of states is shown for non spin-polarized and spin-polarized GGA, and DMFT. For DMFT also the temperature dependence of the feature at the Fermi level (zero energy) is depicted. Here $\beta$ refers to the inverse temperature. The latter contains the local 3d projected, \textit{$d_{z^{2}}$}, \textit{$d_{x^{2}-y^{2}}$}, \textit{$d_{xz}$}/\textit{$d_{yz}$} and \textit{$d_{xy}$}, partial density of states for non spin-polarized and spin-polarized GGA, and DMFT at two different inverse temperatures.

\begin{figure}[!ht]
\begin{center}
\includegraphics[trim=80 40 30 60, clip, width=9cm, scale=0.5]{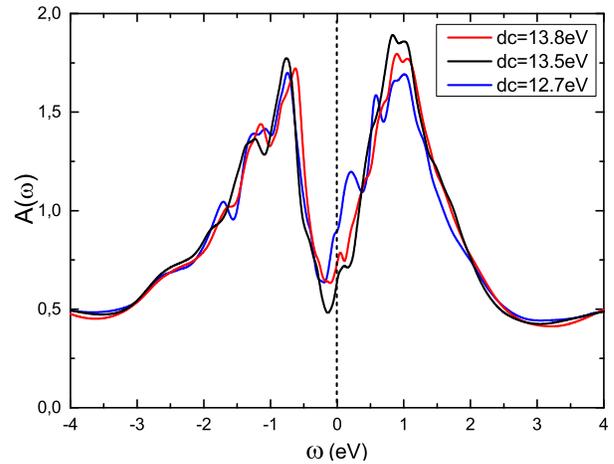}
\end{center}
\caption{The local spin averaged 3d partial density of states within SPTF for different double countings at $\beta=20$~eV$^{-1}$. }
\label{SPTF_dc}
\end{figure}

In order to compare SPTF with CTQMC, the local spin averaged 3d partial density of states is calculated within SPTF for different double countings (see Fig.~\ref{SPTF_dc}) at $\beta=20$~eV$^{-1}$. From a comparison with Fig.~\ref{CTQMCfig2}a (solid black line) it can be observed that for the double countings $13.8$ and $13.5$~eV the height of the feature at the Fermi level is underestimated with respect to the main peak at about $1$~eV. On the other hand for the $12.7$~eV double counting the agreement is very good. There is only a slight mismatch in the position of the feature at the Fermi level. This mismatch will be addressed below in more details. 

\begin{figure}[!ht]
\begin{center}
\includegraphics[trim=80 40 30 60, clip, width=9cm, scale=0.5]{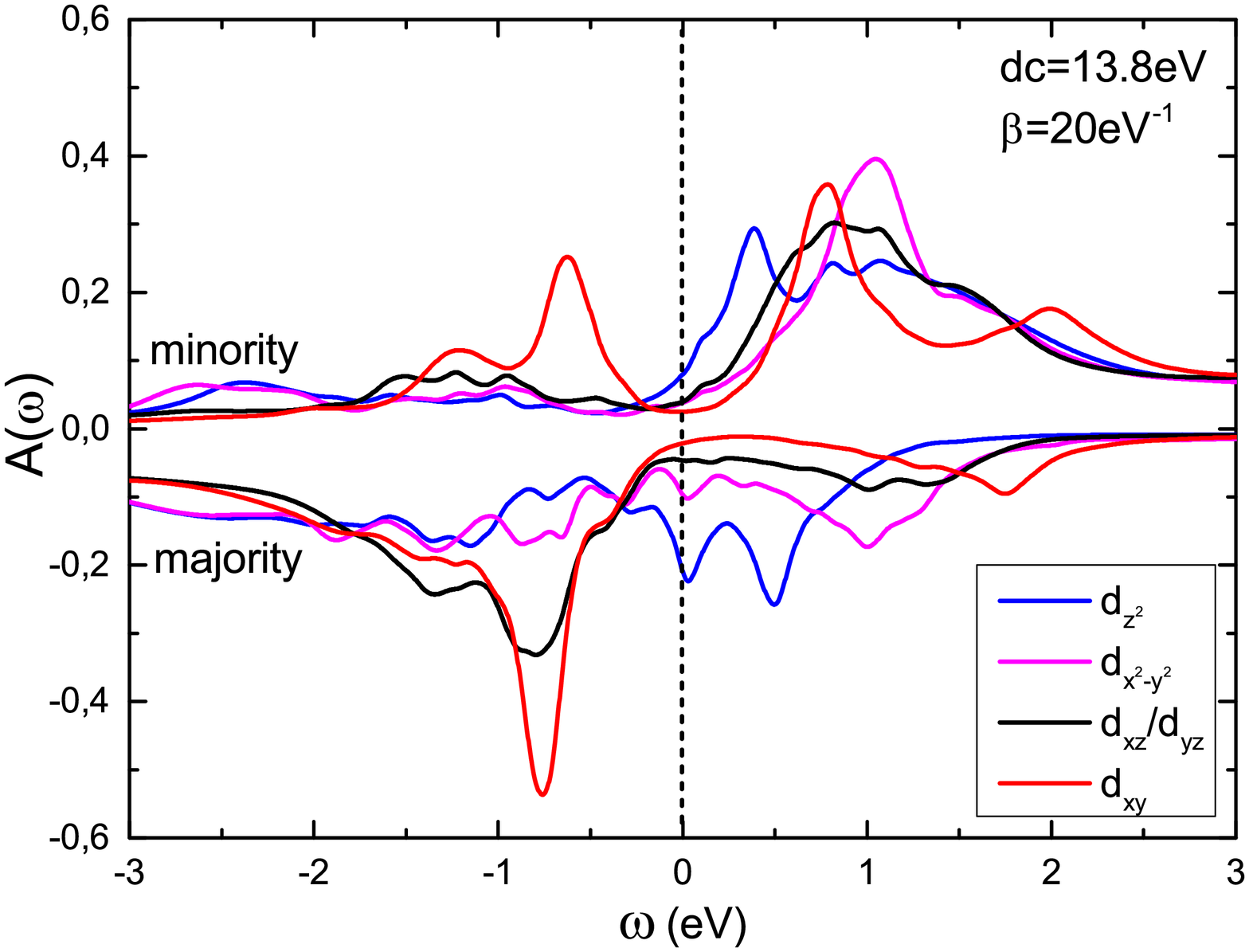}
\includegraphics[trim=80 40 30 60, clip, width=9cm, scale=0.5]{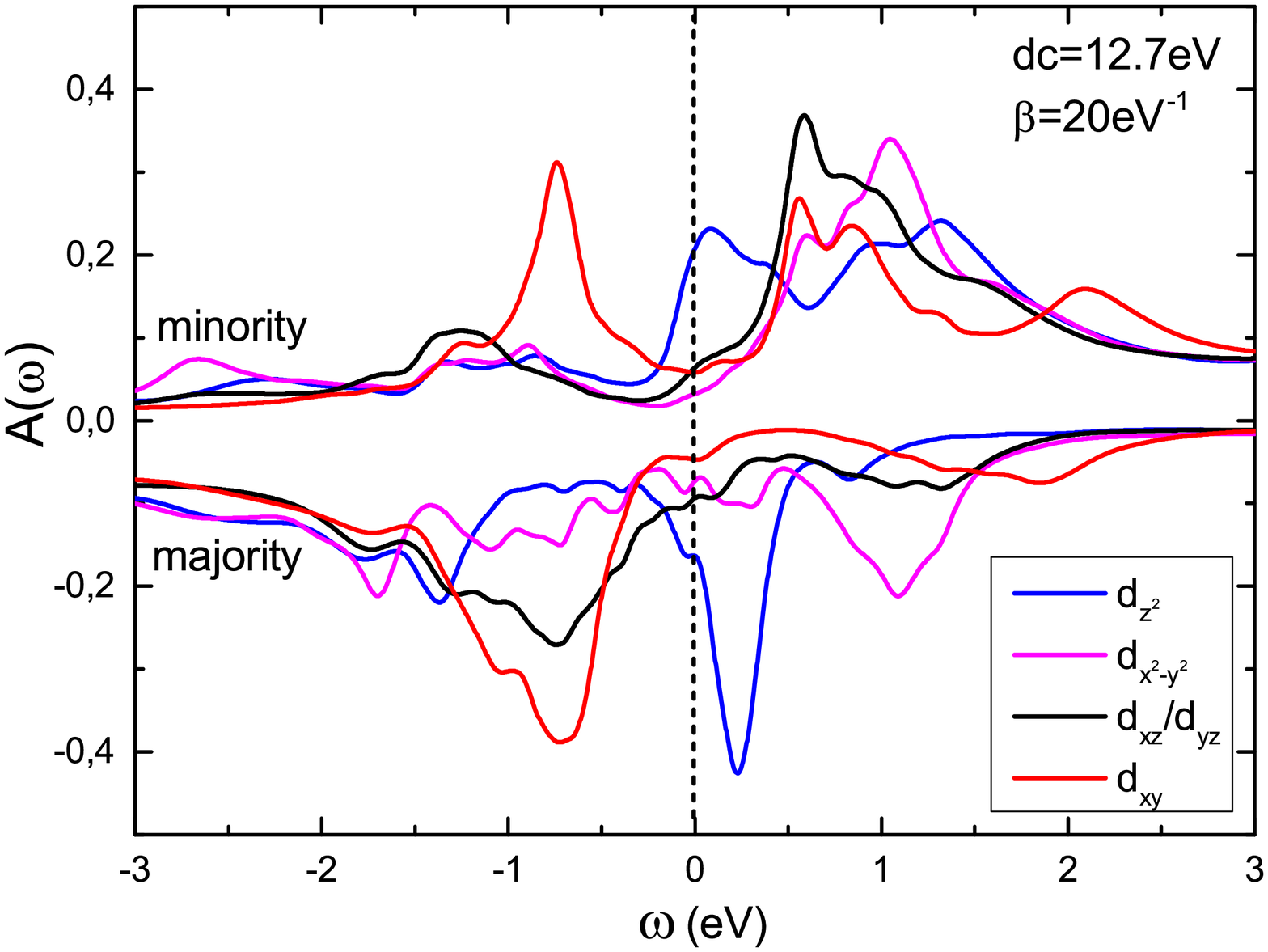}
\end{center}
\caption{The local 3d projected density of states within SPTF for $13.8$~eV (top) and $12.7$~eV (bottom) double counting at $\beta=20$~eV$^{-1}$. Here blue corresponds to \textit{$d_{z^{2}}$}, magenta to \textit{$d_{x^{2}-y^{2}}$}, black to \textit{$d_{xz}$}/\textit{$d_{yz}$} and red to \textit{$d_{xy}$}. }
\label{SPTF_proj}
\end{figure}

It becomes even more clear that SPTF for $12.7$~eV double counting is in good agreement with CTQMC, while that of $13.8$~eV is not, from an inspection of the local 3d projected density of states. In Fig.~\ref{SPTF_proj} the local 3d partial density of states projected on \textit{$d_{z^{2}}$}, \textit{$d_{x^{2}-y^{2}}$}, \textit{$d_{xz}$}/\textit{$d_{yz}$} and \textit{$d_{xy}$} is shown for $13.8$ (top) and $12.7$~eV (bottom) double counting within SPTF for $\beta=20$~eV$^{-1}$. From a comparison with the CTQMC results in Fig.~\ref{CTQMCfig}c (for the same inverse temperature) it is clear that SPTF with $12.7$ double counting is in very good agreement, while that of $13.8$ is not. For example, the calculation for $13.8$ double counting wrongly predicts the main contribution of the feature at the Fermi level to be of majority \textit{$d_{z^{2}}$} type. For the $12.7$ double counting the main contribution is correctly predicted to originate from the minority \textit{$d_{z^{2}}$} channel. However, it should be noted that its contribution is a bit underestimated with respect to CTQMC. Also the majority \textit{$d_{z^{2}}$} state seems to be a bit too close to the Fermi level. Furthermore, the broad features around $-1$ and $+1$~eV are in good agreement with CTQMC.

\begin{figure}[!ht]
\begin{center}
\includegraphics[trim=80 40 30 60, clip, width=9cm, scale=0.5]{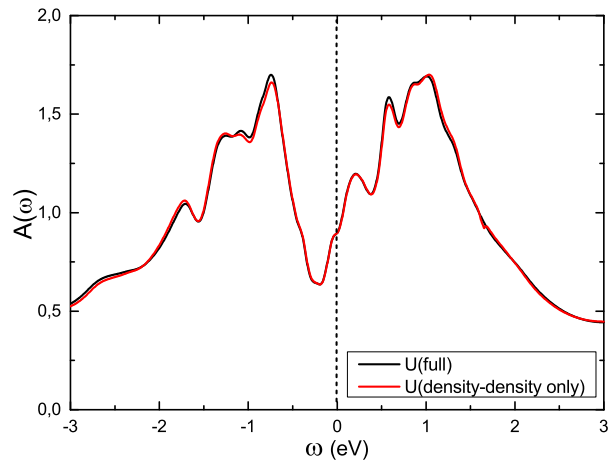}
\end{center}
\caption{The local spin averaged 3d partial density of states within SPTF with full (black) and density-density only Coulomb interaction (red) at $\beta=20$~eV$^{-1}$ for $12.7$~eV double counting. }
\label{SPTF_full_dd}
\end{figure}

As mentioned earlier in CTQMC the treatment of the full Coulomb matrix becomes prohibitively expensive at low temperatures. Therefore, in Ref.~\onlinecite{schuler} only density-density terms of the local Coulomb interaction are considered. It is interesting to see what the influence of this approximation is on the spectrum. For this purpose a SPTF calculation is performed with full and density-density only local Coulomb interaction. In Fig.~\ref{SPTF_full_dd} the local spin averaged 3d partial density of states is depicted for these two calculations, where a $12.7$~eV double counting and $\beta=20$~eV$^{-1}$ was used. From this figure it can be observed that the consideration of the full Coulomb matrix and density-density terms only leads to very similar results. Only the peaks around $-1$ and $+1$~eV are slightly different. 

\begin{figure}[!ht]
\begin{center}
\includegraphics[trim=80 40 30 60, clip, width=9cm, scale=0.5]{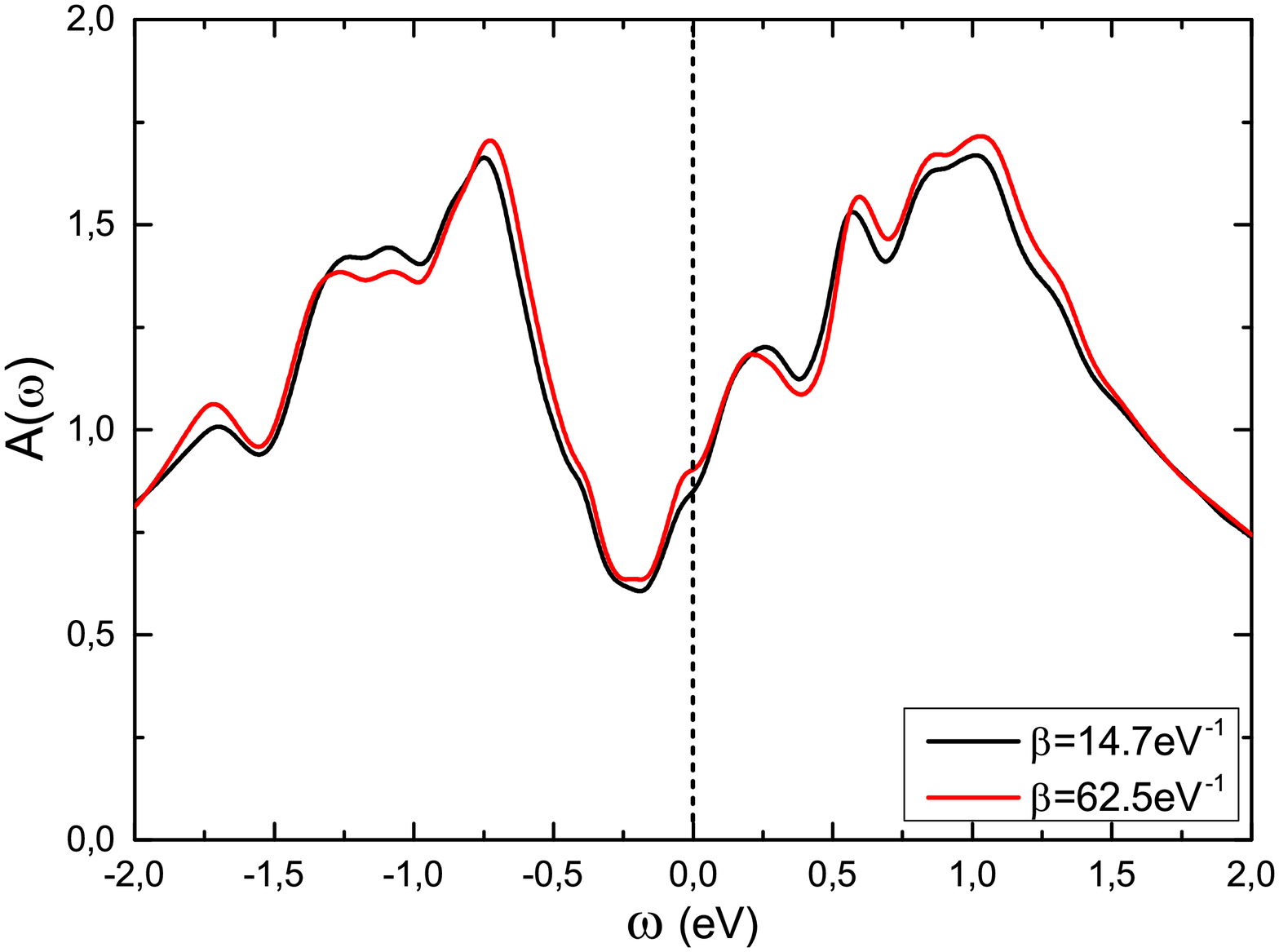}
\end{center}
\caption{The local spin averaged 3d partial density of states within SPTF at two different inverse temperatures $\beta=14.7$~eV$^{-1}$ (black) and $\beta=62.5$~eV$^{-1}$ (red) for $12.7$~eV double counting and full Coulomb interaction. }
\label{SPTF_T}
\end{figure}

It also interesting to investigate the temperature dependence of the spectral feature close to the Fermi level. For the CTQMC calculations this is presented in Fig.~\ref{CTQMCfig2}. Here the feature shifts towards the Fermi level for increasing temperature (decreasing $\beta$). For SPTF the temperature dependent results are shown in Fig.~\ref{SPTF_T}, where for $12.7$~eV double counting the local spin averaged 3d partial density of states is presented for two different inverse temperatures $\beta=14.7$~eV$^{-1}$ (black) and $\beta=62.5$~eV$^{-1}$ (red). From this figure it can be observed that in contrast to the CTQMC results the position of this spectral feature shifts closer to the Fermi level for decreasing temperature. Thus, part of the mismatch in the position of the spectral feature at the Fermi level (between CTQMC and SPTF) is due to the different temperature dependence. Probably the rest of the mismatch is caused by the difference in double counting.

The next step is to perform SPTF calculations for temperatures at which the sharp resonance at the Fermi level is observed experimentally, roughly below 100~K. Therefore, we performed calculations for temperatures as low as 15~K. The results are not shown here, because they are essentially the same as for $\beta=62.5$~eV$^{-1}$ ($T=185.7$~K) shown in Fig.~\ref{SPTF_T}. Thus, the occurence of a sharp resonance at low temperatures is not observed within SPTF. From this result and the good agreement between SPTF and CTQMC at higher temperatures, it can be concluded that the high-energy spectral features, everything except the resonance, are mainly due to the physical processes captured within SPTF.

\begin{figure}[!ht]
\begin{center}
\includegraphics[trim=80 40 30 60, clip, width=9cm, scale=0.5]{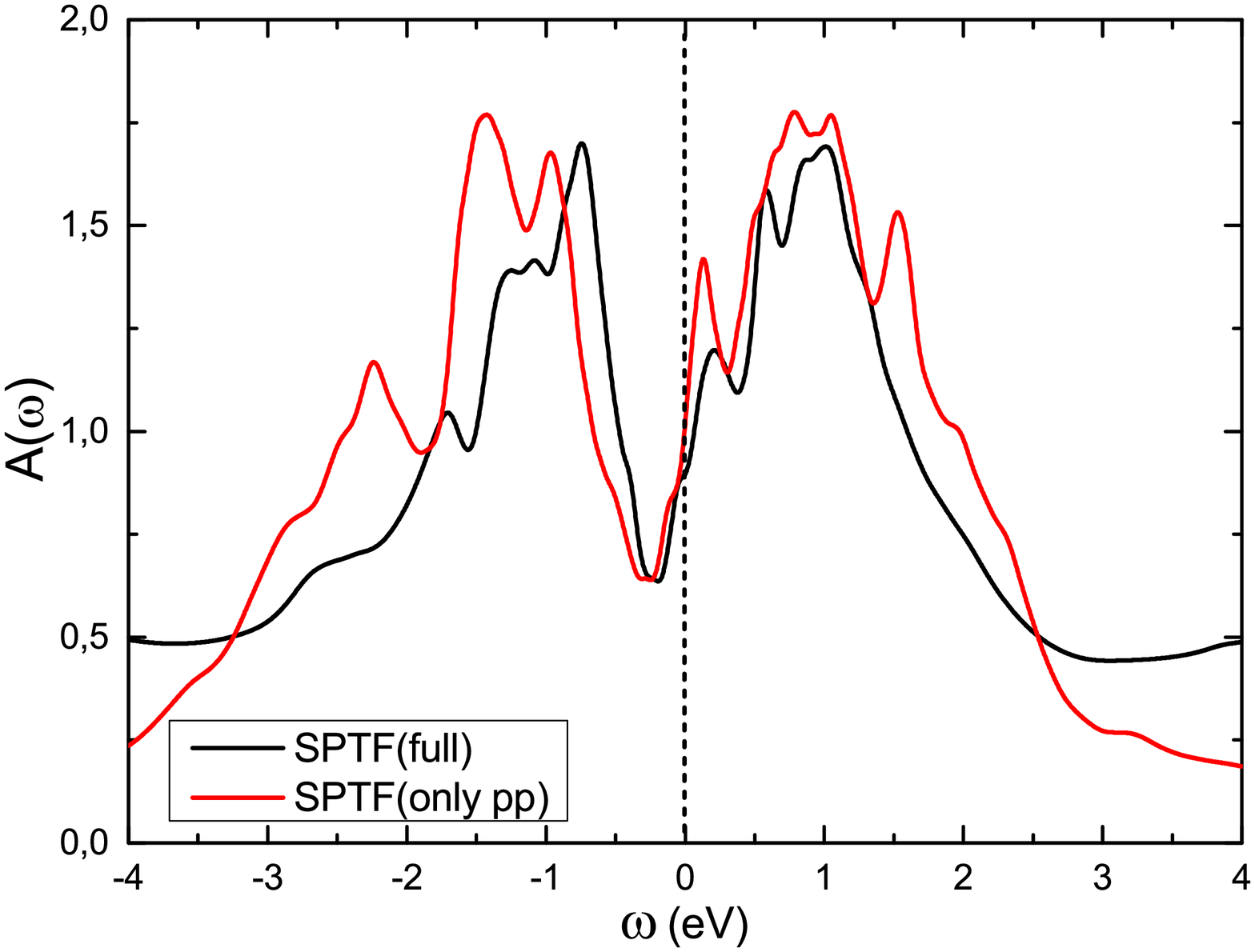}
\includegraphics[trim=80 33 30 60, clip, width=9cm, scale=0.5]{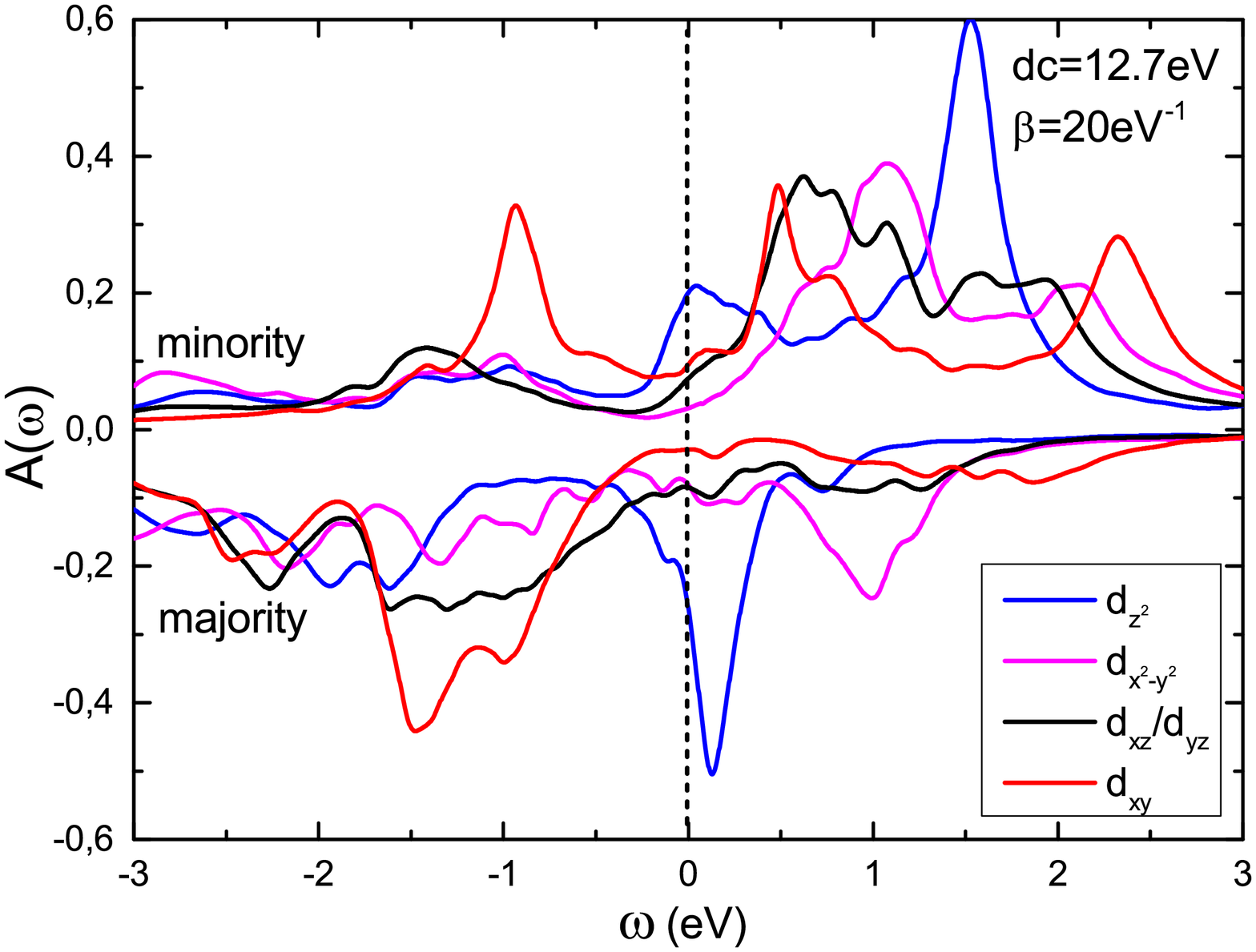}
\end{center}
\caption{In the top figure the local spin averaged 3d partial density of states is presented for full SPTF (black) and SPTF with only particle-particle processes (red). The bottom figure contains the local 3d projected density of states for the SPTF with only particle-particle processes. Here blue corresponds to \textit{$d_{z^{2}}$}, magenta to \textit{$d_{x^{2}-y^{2}}$}, black to \textit{$d_{xz}$}/\textit{$d_{yz}$} and red to \textit{$d_{xy}$}. All these calculations are for $\beta=20$~eV$^{-1}$ and $12.7$~eV double counting. }
\label{SPTF_pponly}
\end{figure}

In order to obtain an even more detailed understanding of which physical processes are dominant for the high-energy spectral features, we performed additional SPTF calculations with particle-hole processes excluded. This means that the contribution of Eq.~\ref{eqsptfph} is not included anymore and only particle-particle processes are considered. In Fig.~\ref{SPTF_pponly} the local spin averaged 3d partial density of states is shown (top figure) of a SPTF calculation with both particle-particle and particle-hole processes included (black) and one with only particle-particle processes considered (red). The bottom figure of Fig.~\ref{SPTF_pponly} contains the local 3d projected density of states for SPTF with only particle-particle processes considered. All these calculations are for $\beta=20$~eV$^{-1}$, $12.7$~eV double counting and full Coulomb interaction. Thus, from a comparison of Figs.~\ref{CTQMCfig2}a and~\ref{CTQMCfig}c with Figs.~\ref{SPTF_dc}, \ref{SPTF_proj} and \ref{SPTF_pponly} it appears that the particle-particle processes provide the main contribution to the high-energy spectral features.

\begin{figure}[!ht]
\begin{center}
\includegraphics[trim=70 30 30 60, clip, width=9cm, scale=0.5]{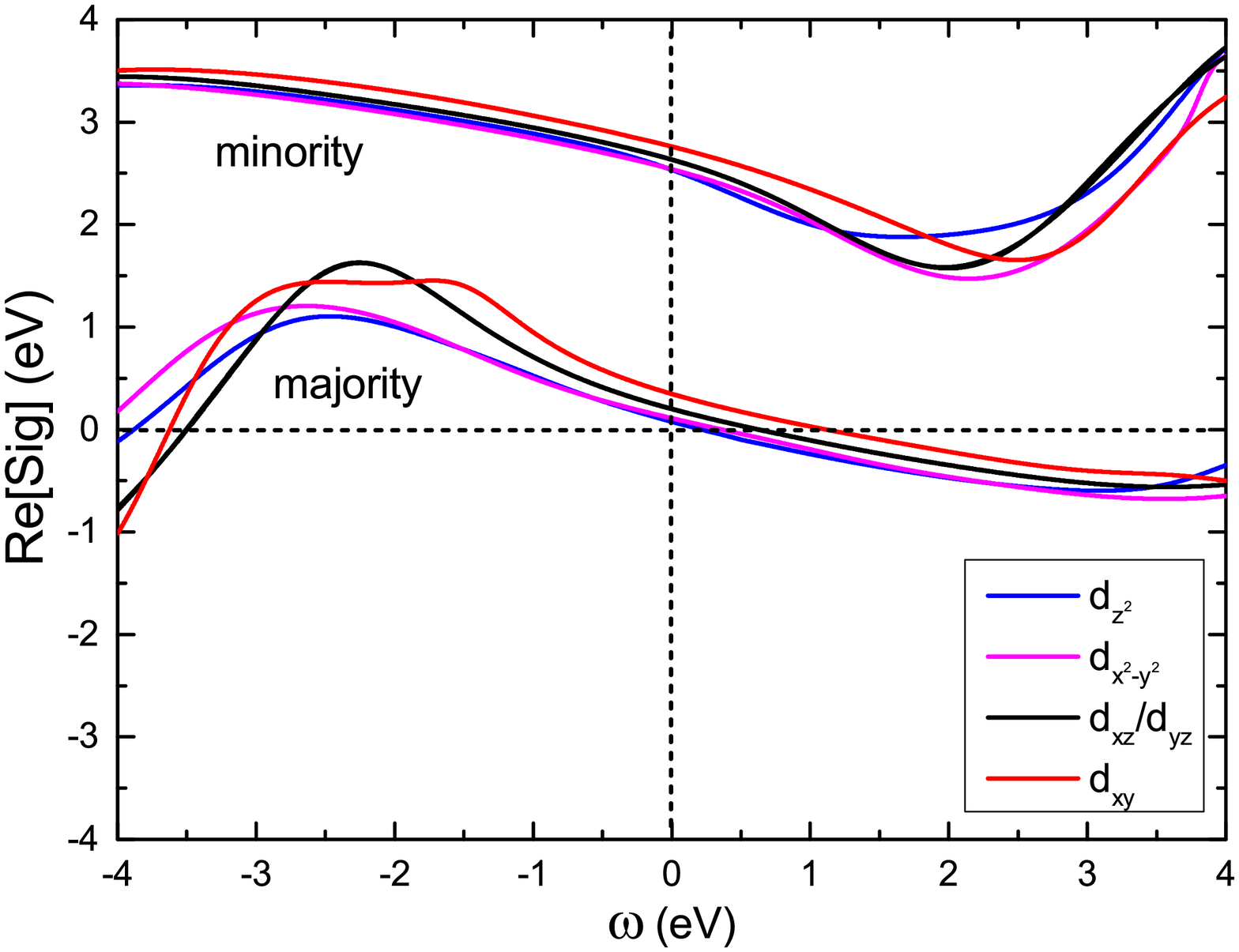}
\includegraphics[trim=70 40 30 60, clip, width=9cm, scale=0.5]{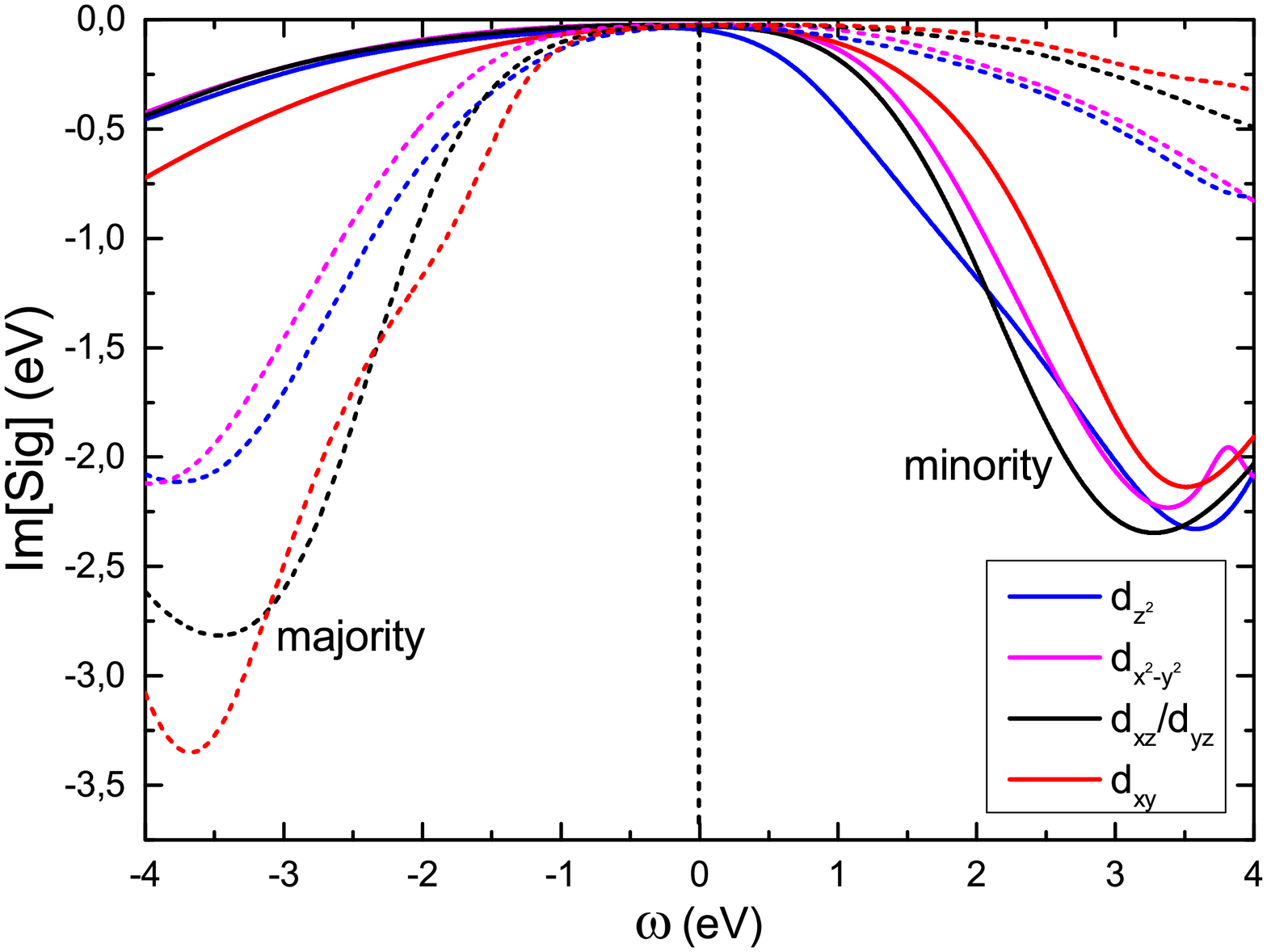}
\end{center}
\caption{The local 3d projected real (top) and imaginary (bottom) part of the self-energy is depicted for a full SPTF calculation with $\beta=20$~eV$^{-1}$, $12.7$~eV double counting and full Coulomb interaction. Here blue corresponds to \textit{$d_{z^{2}}$}, magenta to \textit{$d_{x^{2}-y^{2}}$}, black to \textit{$d_{xz}$}/\textit{$d_{yz}$} and red to \textit{$d_{xy}$}.  }
\label{sigfig}
\end{figure}

Finally, it is also interesting to have a more detailed understanding of how the peaks of the non spin-polarized GGA spectrum (Fig.~\ref{ggafig}) are renormalized due to the inclusion of the many-body processes on the level of SPTF. For this purpose the real and imaginary part of the local 3d projected self-energy is presented in Fig.~\ref{sigfig}. This figure is for a full SPTF calculation with $12.7$~eV double counting, $\beta=20$~eV$^{-1}$ and full Coulomb interaction. From this figure it can be observed for example that the minority \textit{$d_{z^{2}}$} peak at the Fermi level (see the bottom figure of Fig.~\ref{SPTF_proj}) is a renormalization of the \textit{$d_{z^{2}}$} peak at about $-3$~eV of the non spin-polarized GGA spectrum. On the other hand the majority  \textit{$d_{z^{2}}$} peak at about $0.3$~eV is due to a renormalization of the broad peak at about $1$~eV of the non spin-polarized GGA spectrum.

\subsection{NCA}
\begin{figure}[!ht]
\begin{center}
\includegraphics[trim=110 30 30 60, clip, width=9cm, scale=0.5]{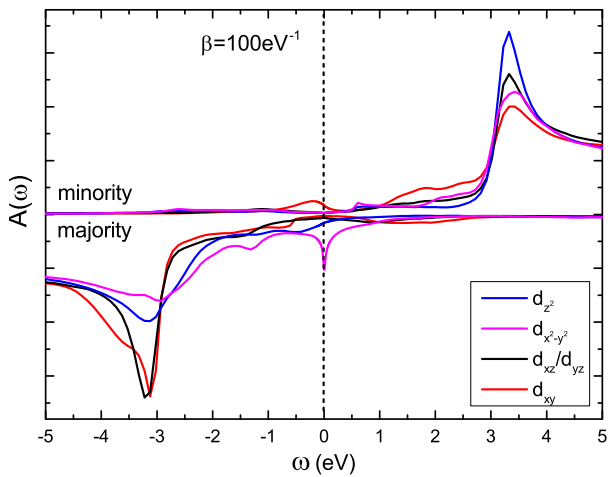}
\end{center}
\caption{The 3d projected partial density of states within NCA for $\beta=100$~eV$^{-1}$. Here blue corresponds to \textit{$d_{z^{2}}$}, magenta to \textit{$d_{x^{2}-y^{2}}$}, black to \textit{$d_{xz}$}/\textit{$d_{yz}$} and red to \textit{$d_{xy}$}.} 
\label{NCAfer}
\end{figure}

We use the NCA scheme to investigate the formation of orbital Kondo-like resonances in the Cr(001) surface at very low temperature. From ferromagnetic NCA calculations with a large spin-splitting of the order of 6~eV (see Fig.~\ref{NCAfer}) we observed a spurious sharp resonance at the Fermi level. In Fig.~\ref{NCAfer} this spurious behaviour can be observed for a one-shot spin-polarized NCA calculation, i.e. the hybridization function and  projected 3d-eigenvalues are obtained from a spin-polarized GGA calculation. The behaviour is spurious, since there is a resonance in the orbitally (and spin) non-degenerate $d_{x^{2}-y^{2}}$ state. Note that the (orbital) Kondo effect is based on a degenerate state. Moreover the self-energy becomes positive, i.e. non-causal. The reason for the occurence of this unphysical behaviour in magnetic NCA calculations is explained in Ref.~\onlinecite{kroh} in terms of missing vertex corrections. Or equivalently, in the presence of a magnetic field, the accidental cancellation at the Kondo temperature of the diverging potential and spin scattering contributions is lifted. Since for non-magnetic NCA this cancellation is complete, there is no unphysical behavior at the Fermi level. Therefore, we restrict ourself in the rest of this work to the non-magnetic Cr(001) case.

From the non-magnetic NCA calculations it appears that the results crucially depend on the behavior of the
hybridization function near the Fermi level. In order to demonstrate this, we used two versions of the non self-consistent calculations, 
which give drastically different spectral functions near the Fermi level. The first one consists of the hybridization function $\Delta$
calculated from the standard non-interacting impurity problem~\cite{dmft1,dmft2}
\begin{equation}
G_{imp}^{-1}(i\omega_{n})=i\omega_{n}-\mu - \Delta(i\omega_{n}).
\label{imp}
\end{equation}
\newline
The second approach is based on the Bethe-lattice approximation~\cite{dmft1} with some adjustable
Bethe-hopping $t_B$
\begin{equation}
\Delta(i\omega_{n})=t^{2}_B G_{imp}(i\omega_{n}).
\label{bethe}
\end{equation}
\newline
In this case we used $t_B$ as a scaling parameter in order to have a similar magnitude for the hybridization function as the ones obtained directly from the impurity GGA calculations.
The main reason to check these models is related to the very different behavior of the hybridization function near the Fermi level in these
two cases: while in the impurity-model we get mainly the
$d_{z^{2}}$ and $d_{x^{2}-y^{2}}$ orbitals at $E_F$, in the Bethe-lattice model the main peaks are related with the
{$d_{xz}$}/{$d_{yz}$} and $d_{xy}$ orbitals, which is clearly seen from the non-magnetic projected partial
density of states (Fig.~\ref{ggafig}).

\begin{figure}[!ht]
\begin{center}
\includegraphics[trim=110 30 30 60, clip, width=9cm, scale=0.5]{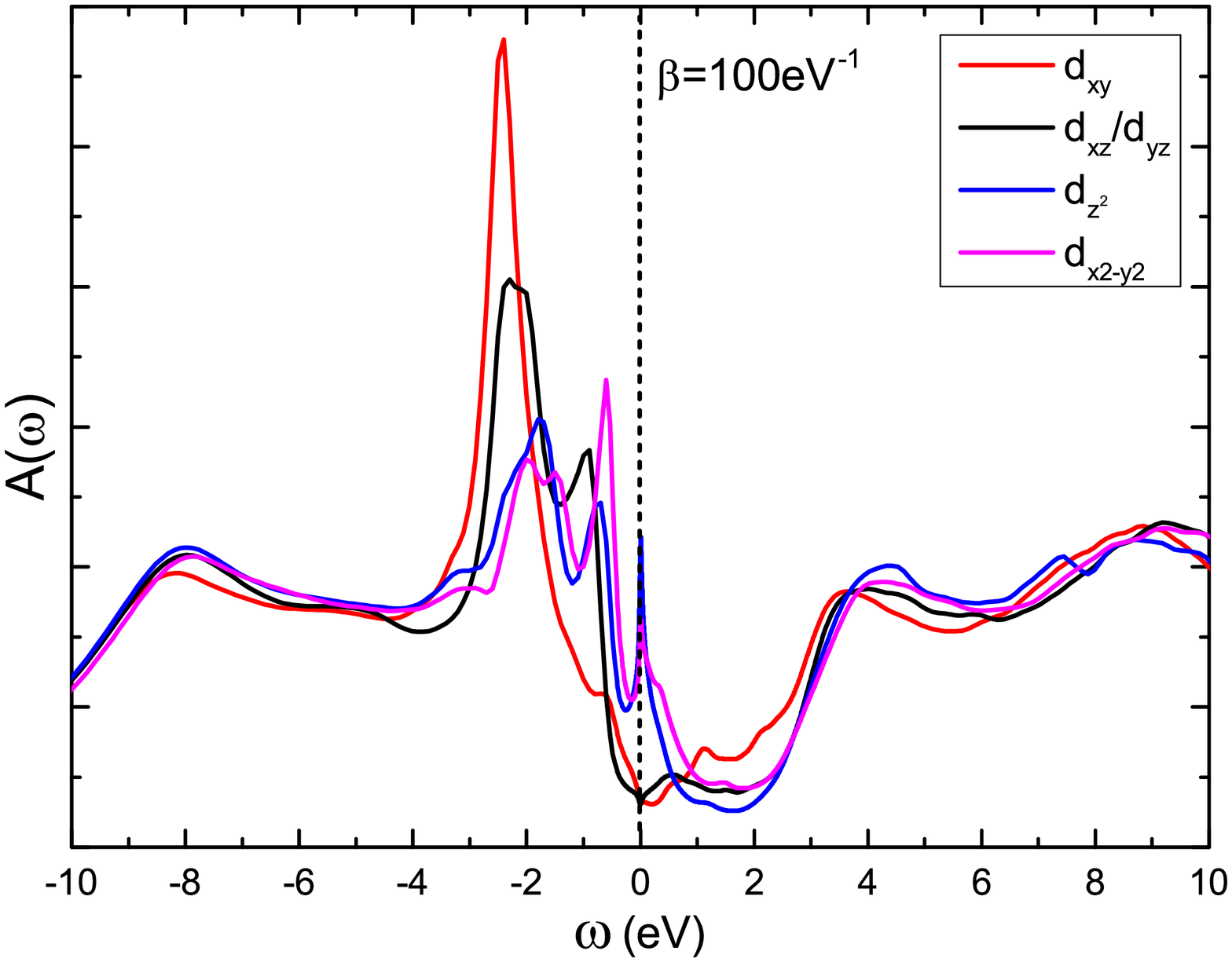}
\includegraphics[trim=110 30 30 60, clip, width=9cm, scale=0.5]{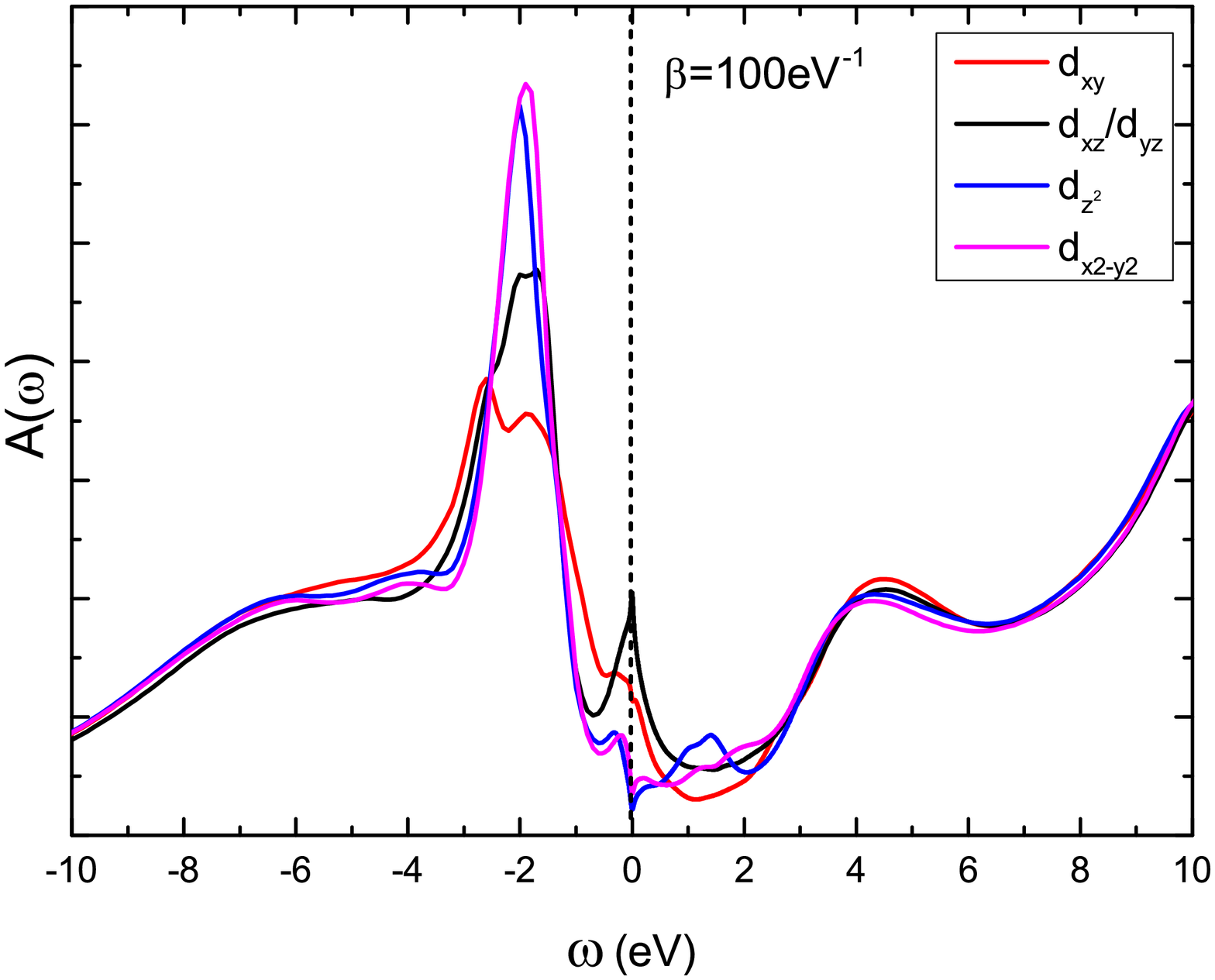}
\end{center}
\caption{The 3d projected partial density of states within NCA for different hybridization functions in non-magnetic Cr(001) for $\beta=100$~eV$^{-1}$. The top figure is for the impurity model and the bottom is for the Bethe-lattice model. Here blue corresponds to \textit{$d_{z^{2}}$}, magenta to \textit{$d_{x^{2}-y^{2}}$}, black to \textit{$d_{xz}$}/\textit{$d_{yz}$} and red to \textit{$d_{xy}$}.}
\label{NCAM}
\end{figure}

Results for the NCA calculations of the non-magnetic Cr(001) surface for both models are presented in Fig.~\ref{NCAM}. It is quite unusual that results are
crucially dependent on the models for the hybridization function: while for the impurity model we have two Kondo-like resonances in the
{$d_{z^{2}}$ and $d_{x^{2}-y^{2}}$ orbitals at $E_F$, for the Bethe-lattice model there is a single broader Kondo resonance in the degenerate {$d_{xz}$}/{$d_{yz}$} orbitals.
The latter corresponds to a strong SU(4) spin-orbit resonance and will reduce to a weaker SU(2) orbital Kondo resonance in the strong magnetic field from the ferromagnetic Cr(001) surface. The former two SU(2) spin resonances in the $d_{z^{2}}$ and $d_{x^{2}-y^{2}}$ orbitals will be killed by a strong magnetic field. Thus, only for the Bethe-lattice model an orbital Kondo resonance could occur in the presence of a strong magnetic field. However, self-consistent spin-polarized NCA calculations for the ferromagnetic state of the Cr(001) surface are needed to unambiguously verify this. Since the present spin-polarized NCA approach suffers from spurious behavior at the Fermi level (Fig.~\ref{NCAfer}), first a thorough investigation of the missing vertex corrections is required to resolve this issue. Such an investigation is out of the scope of this work. It is therefore for future investigations to show a possibility of a self-consistent solution of orbital $d_{xz}$}/{$d_{yz}$} Kondo states in realistic DMFT calculations. We expect that the final hybridization function will be crucially dependent on the starting point which will explain the CTQMC results that used the standard impurity model~\cite{schuler}.

\section{Discussion and conclusion}
In this work we addressed the spectral properties of a Cr(001) surface and in particular the physical origin of the experimentally observed resonance close to the Fermi level at low temperatures. In the literature a single particle $d_{z_{2}}$ surface state renormalized by electron-phonon coupling and the orbital Kondo effect due to the degenerate \textit{$d_{xz}$} and \textit{$d_{yz}$} states are proposed as two possible origins of this resonance. Recent continuous time quantum Monte Carlo calculations within the dynamical mean-field theory already indicated the many body nature of the feature at the Fermi level. However, the precise physical origin of the feature remained unknown. Further, temperatures at which the resonance is experimentally observed could not be reached and only the density-density terms of the Coulomb matrix were considered. 

Therefore, we employed two approximate methods within the dynamical mean-field theory in order to access low temperatures for the full Coulomb matrix and to consider specific physical processes only. First, the spin-polarized T-matrix fluctuation exchange approximation is used, which considers specific scattering processes by treating the onsite Coulomb interaction perturbatively. This method is known to be good for weakly and moderately correlated systems. Second, the non-crossing approximation which is derived in the limit of weak hybridization (strongly correlated systems) and considers Kondo-like processes. 

By using the recent continuous-time quantum Monte Carlo calculations as a benchmark, we found that the high-energy features, everything except the experimentally observed resonance at the Fermi level, of the spectrum is captured within the spin-polarized T-matrix fluctuation exchange approximation. More precisely the particle-particle processes provide the main contribution. The occurence of a resonance even at temperatures as low as 15~K was not observed within this approximation. 

For the non-crossing approximation we found that magnetic calculations lead to a spurious resonance at the Fermi level. Therefore, in order to avoid this unphysical behavior we performed additional non-magnetic calculations. By using two plausible starting hybridization functions, it is shown that the characteristics of the resonance at the Fermi level are crucially dependent on the starting point. For example, in one case a Kondo-like resonance was obtained in the spin degenerate {$d_{z^{2}}$ and $d_{x^{2}-y^{2}}$ orbitals, while in the other case in the spin and orbital degenerate $d_{xz}$/{$d_{yz}$} orbitals. The latter corresponds to a strong SU(4) spin-orbit resonance and will reduce to a weaker SU(2) orbital Kondo resonance in the strong magnetic field from the ferromagnetic Cr(001) surface. The former two SU(2) spin resonances in the $d_{z^{2}}$ and $d_{x^{2}-y^{2}}$ orbitals will be killed by a strong magnetic field. Since we cannot do self-consistent calculations within the present NCA approach for the ferromagnetic state of the Cr(001) surface, it will be very interesting for future investigations to show a possibility of a self-consistent solution of orbital $d_{xz}$}/{$d_{yz}$} Kondo states in realistic DMFT calculations. Before such an investigation can be conducted, a thorough inspection of the missing vertex corrections within the spin-polarized non-crossing approximation is required in order to resolve the spurious behavior at the Fermi level.


\subsection*{Acknowledgements}
We acknowledge support from the Swedish Research Council (VR), eSSENCE, STANDUPP, and the Swedish National Allocations Committee (SNIC/SNAC). The Nederlandse Organisatie voor Wetenschappelijk Onderzoek (NWO) and SURFsara are acknowledged for the usage of the LISA supercomputer and their support. The calculations were also performed on resources provided by the Swedish National Infrastructure for Computing (SNIC) at the National Supercomputer Center (NSC) and the Uppsala Multidisciplinary Center for Advanced Computational Science (UPPMAX). M.I.K. acknowledges a support by European ResearchCouncil (ERC) Grant No. 338957. A.I.L. acknowledges a support from DFG-SFB668. M.K. acknowledges financial support from the Deutsche Forschungsgemeinschaft (DFG) via FOR Grant No. 1162.



\end{document}